%% file: main.tex
\documentclass[a4paper,11pt]{article}
\usepackage{pos}

\input{include/header}
\input{include/FrontMatter}

\begin{document}

\maketitle

\input{content/1-Introduction}
\input{content/2-XmaxMeasurement}
\input{content/3-MassInterpretation}
\input{content/4-PAOAugerComparison}
\input{content/5-Discussion}

\input{include/BackMatter}
\end{document}

%% file: include/header.tex
\usepackage{graphicx}
\usepackage{braket}
\usepackage{multirow}
\usepackage{url}
\usepackage{hvfloat}
\usepackage{commath}
\usepackage{amsmath}
\usepackage{caption}
\usepackage{booktabs}
\usepackage{colortbl}
\usepackage{comment}
\usepackage[export]{adjustbox}
\usepackage{sidecap}
\usepackage{wrapfig}
\usepackage{floatflt} 
\usepackage{enumitem}
\usepackage[nolist,nohyperlinks]{acronym}
\usepackage{blindtext}
\usepackage{lipsum}
\usepackage[left]{lineno}
\usepackage[symbol,bottom]{footmisc}
\renewcommand{\thefootnote}{\fnsymbol{footnote}}

\def\figureautorefname~#1\null{Fig.\,#1\null}
\usepackage{titlesec}
\titlespacing*{\section}
  {0pt}{2.0ex plus 0.5ex minus 0.2ex}{0.8ex plus 0.3ex}
\titlespacing*{\subsection}
  {0pt}{0.8ex plus 0.4ex minus 0.1ex}{0.6ex plus 0.2ex}
  
\DeclareMathSizes{10}{9}{6}{5}

\newlength{\bibitemsep}
\newlength{\bibparskip}
\let\oldthebibliography\thebibliography
\renewcommand\thebibliography[1]{%
  \oldthebibliography{#1}%
  \setlength{\parskip}{\bibitemsep}%
  \setlength{\itemsep}{\bibparskip}%
}

\newcommand{\xmax}{$X_{\text{max}}$}

\newcommand{\xmaxmu}{$\langle X_{\text{max}} \rangle$}

\newcommand{\xmaxsigma}{$\sigma\left( X_{\text{max}} \right)$}

\newcommand{\lnA}{ln$(A)$}
\newcommand{\lnAmu}{$\langle \text{ln}(A) \rangle$}
\newcommand{\VlnA}{$V\left(\text{ln}(A)\right)$}
\newcommand{\lge}{lg$(E/\text{eV})$}

\newcommand{\gcm}{g/cm$^2$}

\acrodef{UHE}{ultra-high-energy}
\acrodefplural{UHE}{ultra-high-energies}
\acrodef{UHECR}{ultra-high-energy cosmic ray}
\acrodefplural{UHECR}{ultra-high-energy cosmic rays}
\acrodef{GMF}{Galactic Magnetic Field}
\acrodef{EGMF}{Extragalactic Magnetic Field}
\acrodef{EB}{Editorial Board}
\acrodefplural{EB}{Editorial Boards}
\acrodef{EAS}{extensive air shower}
\acrodefplural{EAS}{extensive air showers}
\acrodef{SGP}{supergalactic plane}
\acrodef{FD}{Fluorescence Detector}
\acrodef{SD}{Surface Detector}
\acrodefplural{SD}{Surface Detectors}
\acrodef{SBG}{starburst galaxies}
\acrodef{AGN}{active galactic nuclei}
\acrodef{DNN}{Deep Neural Network} 
\acrodefplural{DNN}{Deep Neural Networks}
\acrodef{1DCNN}[1-D CNN]{one dimensional Convolutional Neural Network} 
\acrodefplural{1DCNN}[1-D CNNs]{one dimensional Convolutional Neural Networks} 
\acrodef{CNN}[CNN]{Convolutional Neural Network} 
\acrodefplural{CNN}[CNNs]{Convolutional Neural Networks} 
\acrodef{LSTM}{Long Short-Term Memory}
\acrodefplural{LSTM}[LSTMs]{Long Short-Term Memory networks}
\acrodef{HPC}{High Performance Computing Center}
\acrodef{MURF}{Mines Undergraduate Research Fellowship}
\acrodef{ML}{Machine Learning}
\acrodef{HIM}{Hadronic Interaction Model}
\acrodefplural{HIM}[HIMs]{Hadronic Interaction Models}
\acrodef{GFN}[GFN]{Great Flexible Network}
\acrodef{MSE}[MSE]{Mean Squared Error}
\acrodef{WCD}[WCD]{Water Cherenkov Detector}
\acrodefplural{WCD}[WCDs]{Water Cherenkov Detectors}
\acrodef{SSD}[SSD]{Surface Scintillator Detector}
\acrodefplural{SSD}[SSDs]{Surface Scintillator Detectors}
\acrodef{AERA}[AERA]{Auger Engineering Radio Array}
\acrodef{RD}[RD]{Radio Detector}
\acrodefplural{RD}[RDs]{Radio Detectors}
\acrodef{HEAT}[HEAT]{High Elevation Auger Telescopes}
\acrodef{TS}[TS]{Test Statistic}

%% file: include/FrontMatter.tex
\title{Measurement and Interpretation of UHECR Mass Composition at the Pierre Auger Observatory}
\ShortTitle{Mass Composition at the Pierre Auger Observatory}

\author*[a]{Eric Mayotte}
\onbehalf{for the Pierre Auger Collaboration$^b$}

\affiliation[a]{Department of Physics, Colorado School of Mines, 1523 Illinois St., Golden CO, USA}
\affiliation[b]{Observatorio Pierre Auger, Av.\ San Mart{\'\i}n Norte 304, 5613 Malarg\"ue, Argentina\\
Full author list: \normalfont{\url{https://www.auger.org/archive/authors\_icrc\_2025.html}}}

\emailAdd{spokespersons@auger.org}

\abstract{The Pierre Auger Observatory has driven the field of ultra-high-energy cosmic ray (UHECR) physics, producing several groundbreaking observations over the last 20 years. One of the most striking findings has been the complex evolution of UHECR mass composition, as revealed by detailed analyses of observables such as the depth of shower maximum ($X_{\rm max}$) and the muon content of showers. As more data are collected and sophisticated analyses are undertaken, not only are new fine details emerging, but the general picture of UHECR mass composition is becoming increasingly robust. This contribution presents recent results on the mass composition of UHECRs derived from surface, fluorescence, and radio detectors. Together with other key findings from the Observatory, these results converge to present a coherent picture of UHECR mass composition, effectively ruling out proton dominance and challenging the interpretation of the observed flux features as purely proton-induced propagation effects. To finish the contribution, we compare the $X_{\rm max}$ data from the southern and northern equatorial bands of the exposure of the Pierre Auger Observatory fluorescence detector to evaluate the possibility of changes in composition as a function of declination.}

\ConferenceLogo{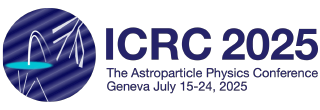}

\FullConference{39th International Cosmic Ray Conference (ICRC2025)\\
 15–24 July 2025\\
Geneva, Switzerland\\}

%% file: content/1-Introduction.tex
\section*{Phase I and Phase II of the Pierre Auger Observatory}

The Pierre Auger Observatory, located near the town of Malarg\"{u}e, Mendoza, Argentina, sits at approximately 35.25$^\circ$\,S, 69.3$^\circ$\,W and has been observing \ac{UHECR} air showers since January 1, 2004.
Since its full completion in June 2008, it has been conducting precision measurements with 1660 \ac{WCD} stations, the \ac{SD}, and 27 fluorescence telescopes, the \ac{FD}~\cite{PierreAuger:2015eyc}.
The Observatory operated in this configuration until 2021, when the deployment of its major upgrade, AugerPrime~\cite{PierreAuger:2016qzd}, began.

The now-completed upgrade to the Observatory significantly increases statistics for mass composition studies at the highest energies through the addition of \acp{SSD} and \acp{RD}. 
The \acp{SSD} complement the WCDs to provide enhanced electromagnetic-muonic shower component separation up to a zenith angle of $60^\circ$. 
The \acp{RD} extend this sensitivity above $60^\circ$ by measuring the electromagnetic component, while the \acp{WCD} measure the muons, which alone survive to the ground at high inclinations.
A small PMT has been added in each \ac{WCD} to enhance their dynamic range. 
Finally, \ac{SD} electronics have been upgraded, providing improved timing resolution.
The data collected before this upgrade constitute the Phase I dataset, while the data collected with the upgraded detector represent the Phase II dataset~\cite{Castellina:2019irv}.

Herein, we present an overview of the current understanding of \ac{UHECR} mass composition, based on data collected during Phase I of the Observatory.
These include newly finalized \ac{FD} measurements using the full Phase I dataset~\cite{Auger:2025InPrep} and new data from a Universality-based \ac{SD} reconstruction~\cite{ICRC25:Universality}.
When these are added to the published measurements from the \ac{SD} via neural networks~\cite{PierreAuger:2024nzw} and the \ac{AERA}~\cite{PierreAuger:2023rgk}, a detailed and consistent picture of the composition of \ac{UHECR} comes into view.
Finally, using the new higher statistics available in the \ac{FD}, the composition of \acp{UHECR} in the southern and northern parts of the Observatory's exposure are directly compared to check for a declination dependence. 

%% file: content/2-XmaxMeasurement.tex
\section*{Measurements of the Depth of Shower Maximum}

When a primary UHECR arrives at Earth and strikes a nucleus in the atmosphere, it triggers a complex cascade of secondary particles known as an \ac{EAS}.
Early in the \ac{EAS}, each collision of a secondary produces further secondaries, primarily pions ($\pi^\pm$, $\pi^0$) and kaons.
The neutral $\pi^0$ quickly decays into pairs of photons, triggering electromagnetic sub-showers through processes like pair production and bremsstrahlung. 
The charged $\pi^\pm$ and kaons either interact further, continuing the process, or decay to muons, creating the muonic component of the shower.

The evolution of the number of particles as a function of atmospheric depth (typically expressed in \gcm{}) is referred to as the longitudinal profile of shower development.
Early in the shower, the particle count grows rapidly as interactions with the atmosphere efficiently convert the initial kinetic energy of the primary into an exponential growth of secondary particles.
This growth continues until the energy per particle falls below the threshold required to readily produce additional secondaries. 
At this critical point, the number of particles reaches its peak, the \emph{Shower Maximum}. 
Beyond this point, secondary particle production is suppressed, and the shower progressively attenuates due to energy losses through ionization, bremsstrahlung, and decay processes, leading to a gradual reduction in particle count as the \ac{EAS} penetrates deeper into the atmosphere.

The atmospheric depth at which this shower maximum occurs is called the \emph{Depth of Shower Maximum} (\xmax{}).
\xmax{} is an important observable as it correlates strongly with the mass number, $A$, of a primary. 
As was laid out in~\cite{kampert2012measurements}, this strong correlation arises because, to first order, the length growth phase of the shower depends on the initial energy per nucleon of the primary.
For a given primary energy, \emph{light} primaries with lower $A$ values will have higher per nucleon energies, leading to, on average, longer growth phases and higher \xmax{} values.
\emph{Heavy} primaries with higher $A$ values, on the other hand, will, on average, have a shorter growth phase and lower \xmax{} values.
However, due to nuclear modifications in nucleus-nucleus interactions, the superposition model is not exact. 
Numerically, the difference, $\langle X_{\rm max}(p,E_p)- \langle X_{\rm max}(A, E_p*A)$, can reach up to 10\,\gcm{} for iron-induced showers simulated with EPOS-LHC.
It is also critical to note that the low particle count in the early stages of the \ac{EAS} leads to sizable statistical fluctuations in the first few interactions, which carry into \xmax{}.
This effect is fully present in light primaries, resulting in high fluctuations in \xmax{}.
In contrast, at first interaction, heavy nuclei create up to $A$ sub-showers which average out this effect, resulting in lower event-to-event variation in \xmax{}.
Here, and more so than for \xmaxmu{}, variations in the nuclear fragmentation of a primary increase shower-to-shower fluctuations over what would be expected from the superposition model alone~\cite{JEngel:1992vf, Kalmykov:1993qe}. 
As a result $\sigma(X_{\rm max})_A > \sigma(X_{\rm max})_p / \sqrt{A}$.

Except in extreme cases, \xmax{} can not be used to estimate the mass of the primary in a single shower due to these shower-by-shower fluctuations.
Instead, typically, many \xmax{} measurements for events in a small energy range are gathered together into distributions of \xmax{}, whose statistical properties, typically their first and second moments, can then be used to estimate the average mass composition of measured \acp{UHECR} in that energy range.
Since the last reporting of the combined \xmax{} measurements from the Observatory~\cite{PierreAuger:2023bfx}, the Phase I \ac{FD} hybrid \xmax{} analysis has been finalized~\cite{Auger:2025InPrep}, and a Universality-based \ac{SD} \xmax{} reconstruction has been carried out~\cite{ICRC25:Universality}. 
For details on the Universality reconstruction, the reader is encouraged to refer to the separate proceedings, which provide a comprehensive outline of the method and results~\cite{ICRC25:Universality}.
The new \ac{FD} hybrid measurement, described in~\cite{PierreAuger:2023kjt}, substantially improves both statistics and analysis. 
The Phase I hybrid dataset contains more than double the number of events of the 2014 \xmax{} PRD, allowing for the addition of new energy bins at high and low energies.

\begin{figure}[!b]
\vspace{-2mm}
  \centering
  \hfill
  \includegraphics[width=0.45\textwidth]{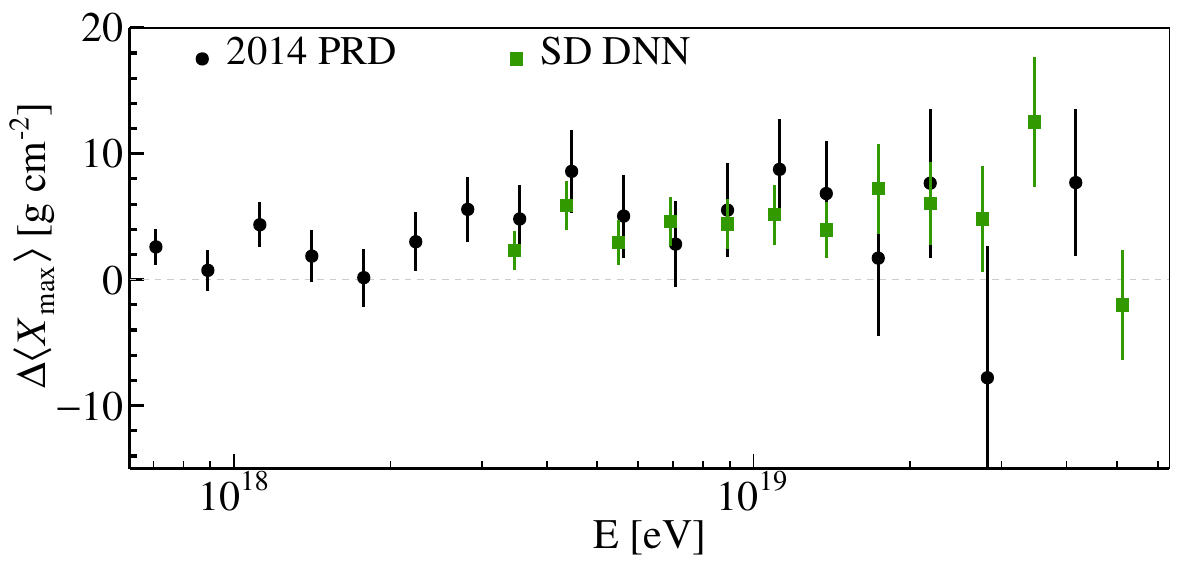}%
  \hfill
  \includegraphics[width=0.45\textwidth]{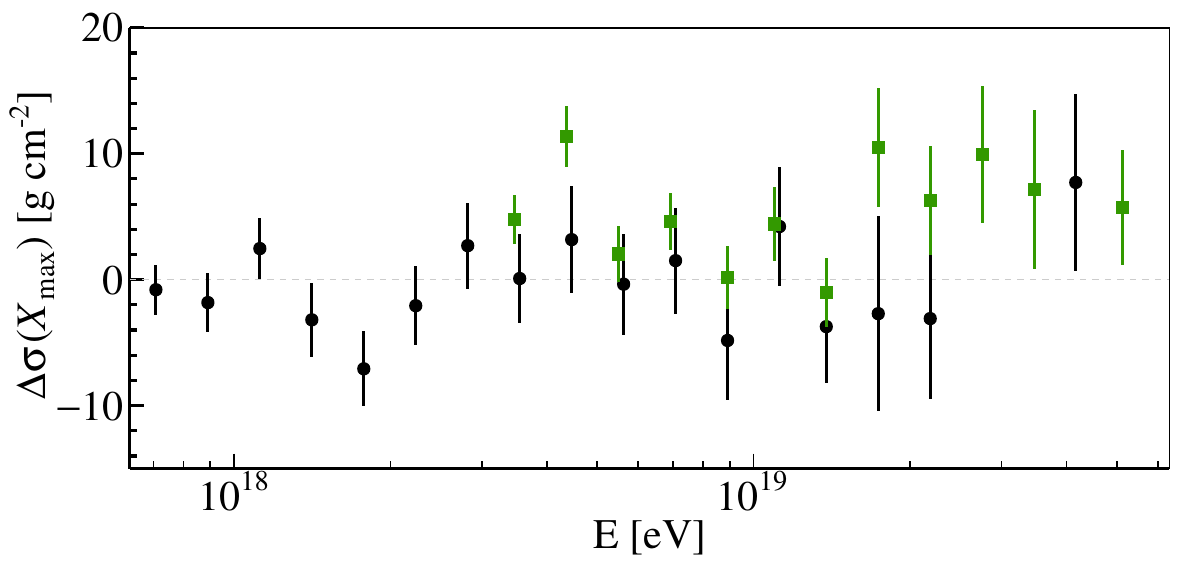}
  \hspace{3mm}
  \caption{\footnotesize Comparison of the Phase I FD \xmax{} measurements~\cite{Auger:2025InPrep} to those published in the 2014 \ac{FD} hybrid and the 2024 \ac{SD} DNN analyses. Left: first moment, $\langle X _{\rm max}^{\rm new} \rangle - \langle X_{\rm max}^{\rm other}\rangle$. Right: second moment $\sigma (X_{\rm max}^{\rm new}) - \sigma (X_{\rm max}^{\rm other})$.}
  \label{fig:XmaxMomentResiduals}
\end{figure}

In comparison to the 2014 result, the Phase I FD Hybrid analysis uses a more accurate parameterization of atmospheric aerosols~\cite{PierreAuger:2023nbk}, an improved method of fitting the shower profile~\cite{PierreAuger:2023att}, and new parameterizations of the fiducial field of view cuts, acceptance, bias, resolution, and systematics~\cite{PierreAuger:2023kjt}. 
These changes have resulted in an, on average, $\sim5$\,\gcm{} shift in reconstructed \xmax{} to higher values.
This difference can be seen in the black points in \autoref{fig:XmaxMomentResiduals} on the left, which shows the mean difference between the new \ac{FD} hybrid \xmax{} reconstruction, the 2014 \ac{FD} \xmax{} measurement~\cite{PierreAuger:2014sui}, and the \ac{SD} DNN \xmax{} reconstruction~\cite{PierreAuger:2024nzw}.
Because the DNN-based \ac{SD} \xmax{} analysis was calibrated using the old \ac{FD} hybrid reconstruction to remove the effects of the hadronic interaction model it was trained on, it also inherited the old \ac{FD} \xmax{} scale and the $\sim 5$\,\gcm{} difference.
Notably for \xmaxsigma{}, no difference is seen between the 2014 \ac{FD} measurement and the new results; however, there is an apparent difference with the \ac{SD} DNN.
The \ac{FD} result consistency rules out changes in \ac{FD} analysis as a cause.
Instead, the difference is likely due to the large \ac{SD} DNN systematic uncertainties in \xmaxsigma{} and residual model or methodological dependencies.

\begin{figure}[!t]
  \vspace{-2mm}
  \centering
  \hfill
  \includegraphics[width=0.45\textwidth]{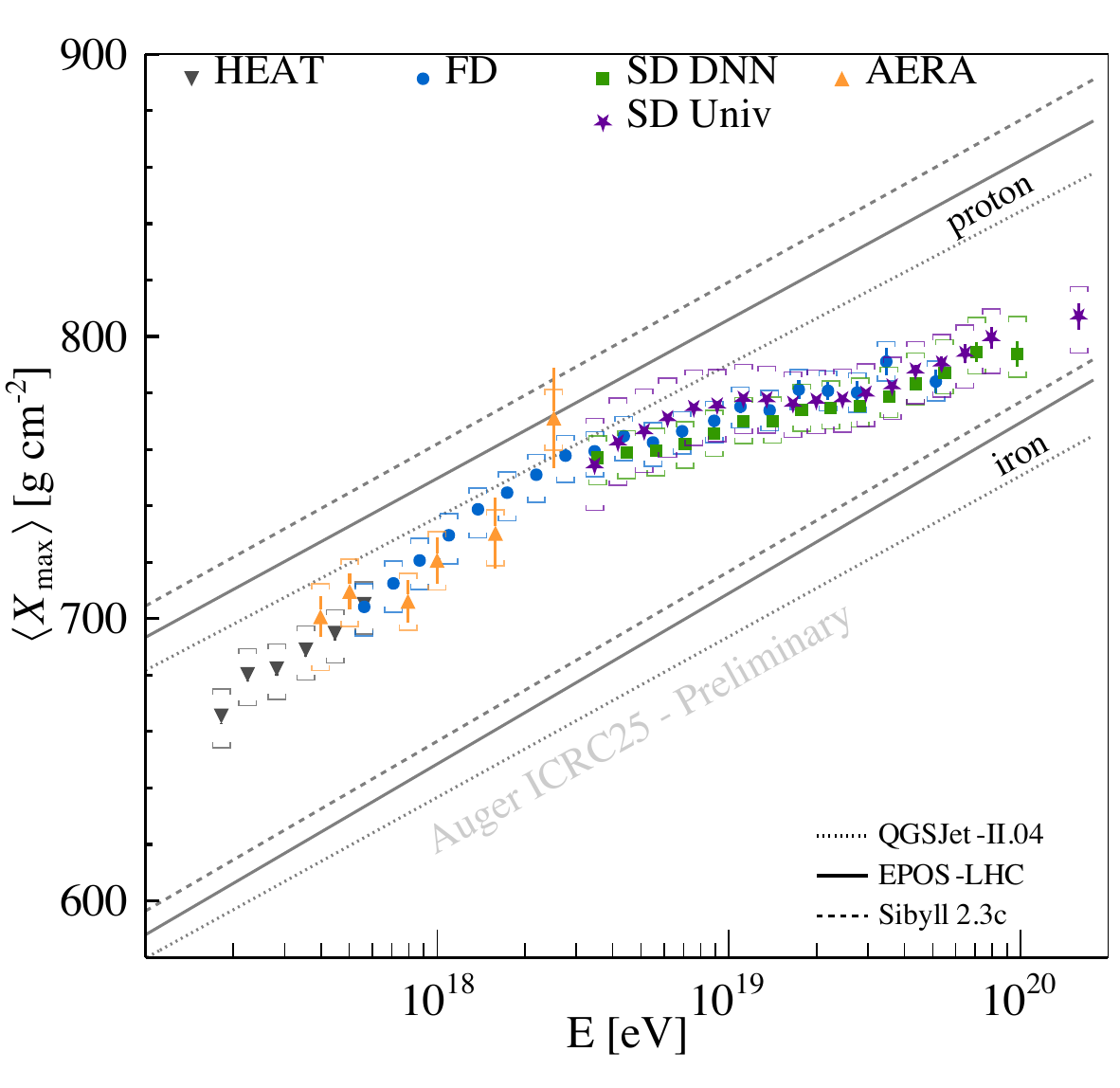}%
  \hfill
  \includegraphics[width=0.45\textwidth]{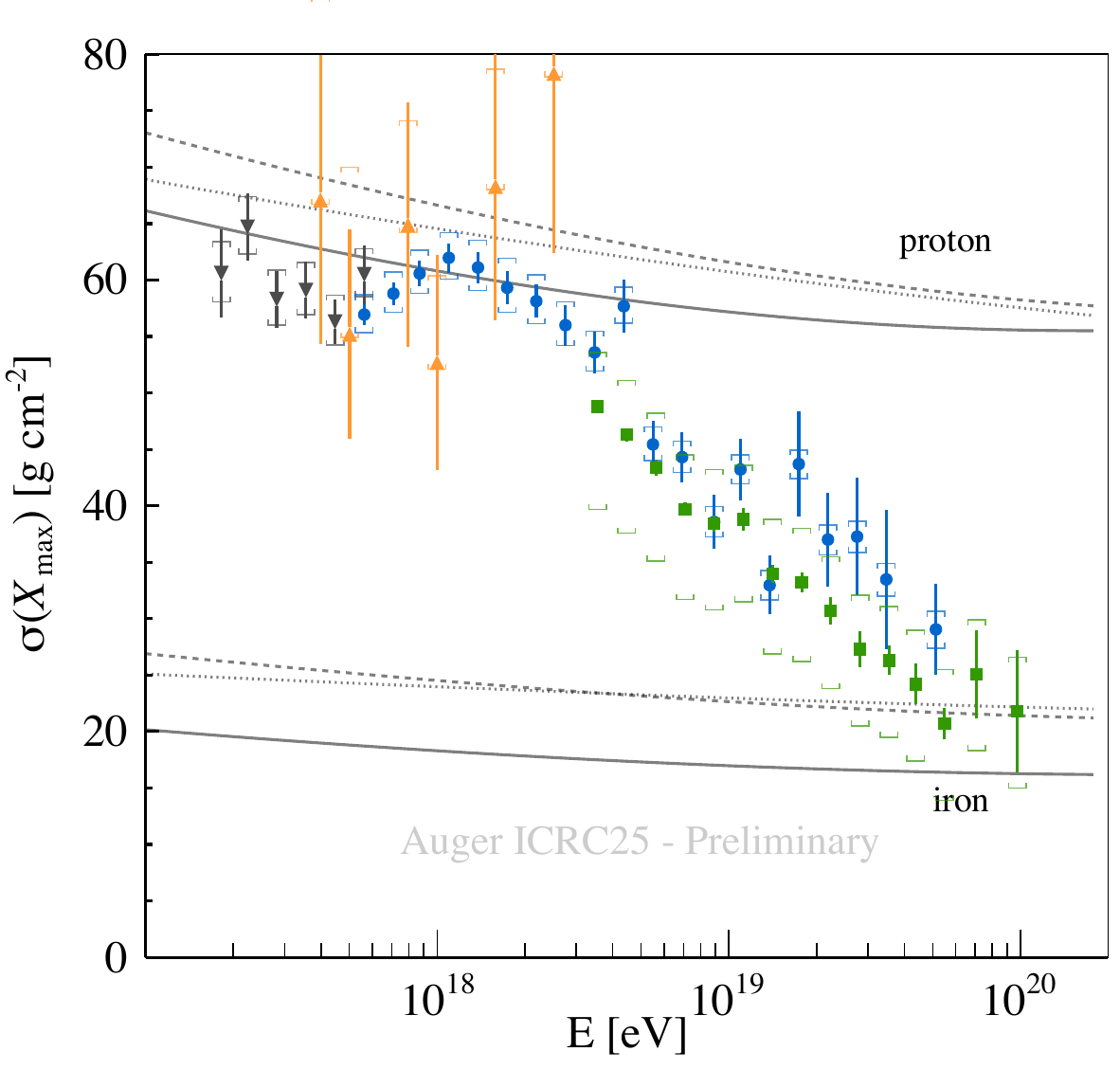}
  \hfill
  \vspace{-2mm}
  \caption{
  \footnotesize The first (left) and second (right) moments of \xmax{} distributions measured with the FD~\cite{Auger:2025InPrep}, the SD~\cite{PierreAuger:2024nzw, ICRC25:Universality}, AERA~\cite{PierreAuger:2023rgk} during Phase I. 
  Preliminary measurements from HEAT~\cite{Bellido:2017cgf} are also shown. 
  }
  \label{fig:XmaxMoments}
\end{figure}

The most up-to-date summary of the latest \xmax{} measurements made using \ac{FD} Hybrid~\cite{Auger:2025InPrep}, SD~\cite{PierreAuger:2024nzw,ICRC25:Universality}, and AERA~\cite{PierreAuger:2023rgk,PierreAuger:2023lkx} data are shown together in \autoref{fig:XmaxMoments}.
Preliminary measurements from \ac{HEAT}~\cite{Bellido:2017cgf} are also included.
Despite progressive updates to reconstruction methods that introduce varying \xmax{} scales in the \ac{FD}, \ac{HEAT}, \ac{AERA}, and \ac{SD} measurements, the \xmaxmu{} values obtained show outstanding consistency at all energies.
\xmaxsigma{} is more complex, and the 18.6–18.7 \lge{} energy bin of the hybrid data in particular stands out.
This energy bin contains a deep outlier event. 
Its removal would decrease \xmaxsigma{}{} by $\sim 2$ \,\gcm{} bringing the moment closer to the overall trend  (see~\cite{Auger:2025InPrep} for details). 
Additionally, \xmaxsigma{} for SD Universality has been omitted due to ongoing work correcting for its resolution (see~\cite{ICRC25:Universality}).

%% file: content/3-MassInterpretation.tex
\section*{The Overall Picture of Mass Composition}
\label{sec:interpretation}

Much information on the mass composition of \acp{UHECR} can be gleaned directly from the moments of \xmax{} shown in \autoref{fig:XmaxMoments}.
That said, it is useful to process the \xmax{} moments and distributions further to extract a fuller picture.
Two practical approaches are to transform the moments of \xmax{} into moments of \lnA{} (the logarithm of the primary mass)~\cite{PierreAuger:2013xim} and to fit the fractional contribution of different mass groups to the flux using the \xmax{} distributions measured at each energy~\cite{PierreAuger:2014gko}.

\begin{SCfigure}[][!t]
    \vspace{-1mm}
    \centering
    \includegraphics[width = 0.36\textwidth]{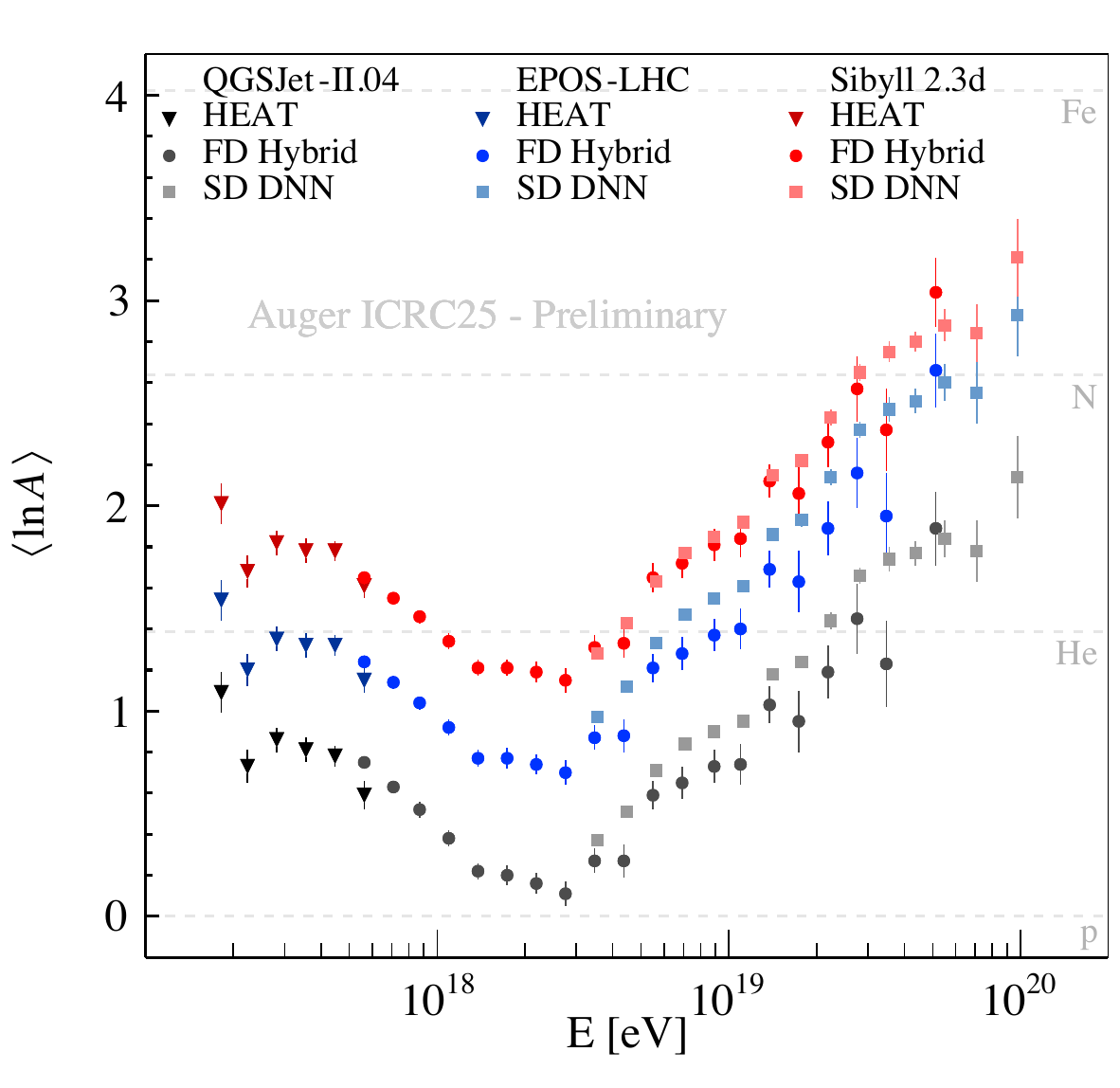}
    \includegraphics[width = 0.36\textwidth]{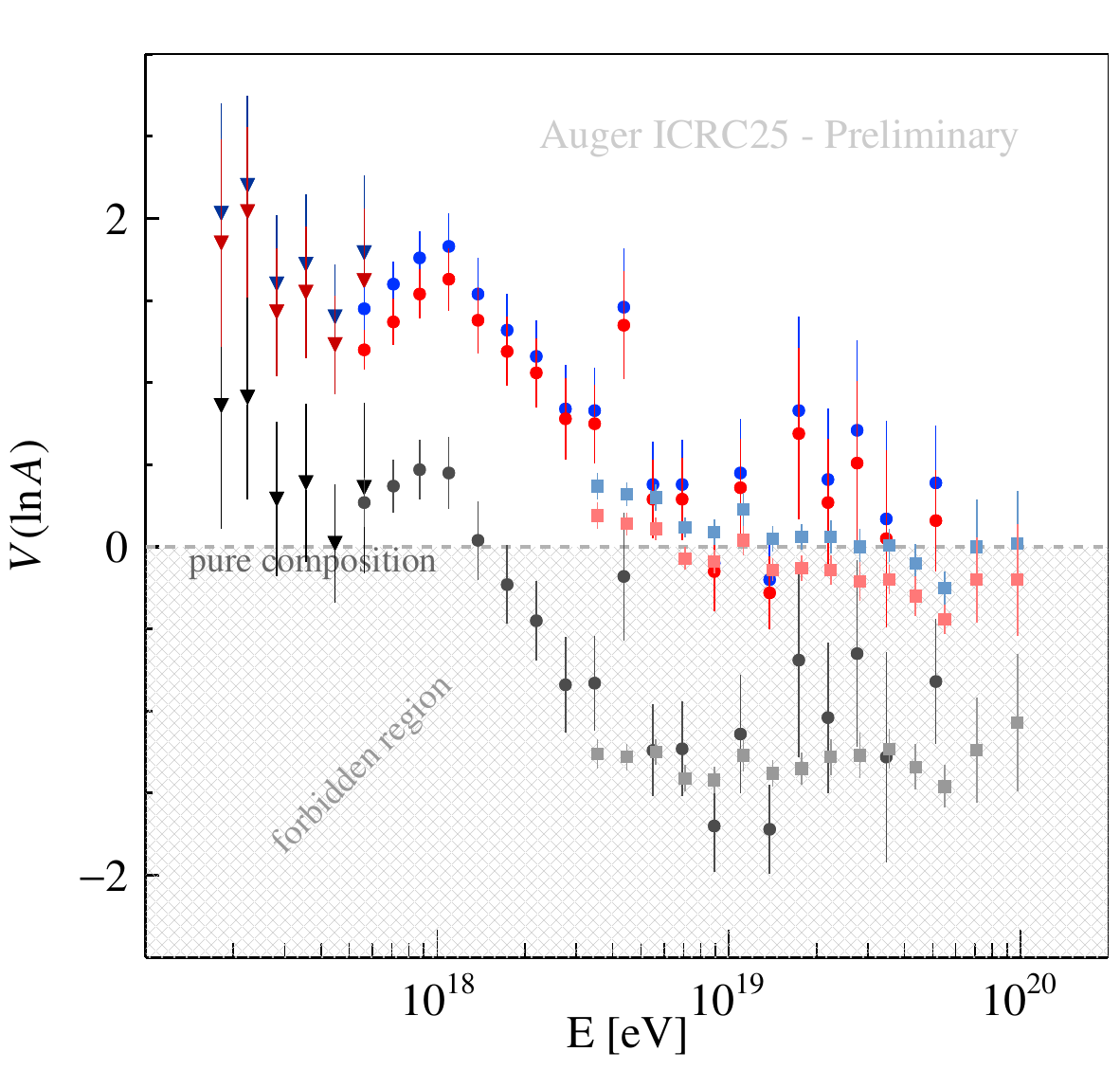}
    \caption{The first (left) and second (right) moments of $\ln{A}$ distributions derived from the FD~\cite{Auger:2025InPrep} and SD Phase I~\cite{PierreAuger:2024nzw} \xmax{} moments in \autoref{fig:XmaxMoments} using QGSJet-II.04 (grey)~\cite{Ostapchenko:2010vb}, EPOS-LHC (blue)~\cite{Pierog:2017awp}, and Sibyll 2.3d (red)~\cite{Riehn:2017mfm}.}
    \label{fig:lnAMoments}
\end{SCfigure}

From~\cite{PierreAuger:2013xim}, the conversion of \xmaxmu{} and \xmaxsigma{} into \lnAmu{} and \VlnA{} uses \acp{HIM} to set the \xmax{} scale and parametrize its fluctuations as
\begin{equation}
    \langle \text{ln}(A) \rangle = \frac{\langle X_{\rm max} \rangle - \langle X_{\rm max} \rangle_p}{f_E},
\end{equation}
where $\langle X_{\rm max} \rangle_p$ is the mean \xmax{} for protons at the relevant energy in the chosen \ac{HIM}, and $f_E = (\langle X_{\rm max} \rangle_{Fe} - \langle X_{\rm max} \rangle_{p})/\ln(56)$. 
\VlnA{}, in turn, is calculated as
\begin{equation}
    V\left(\text{ln}(A)\right) = \frac{\sigma^2\left( X_{\text{max}} \right) - \sigma^2_{sh}\left( \langle \text{ln}(A) \rangle \right)}{b\,\sigma^2_p - {f_E}^2},
\end{equation}
where $\sigma^2_{sh}\left( \langle \text{ln}(A) \rangle \right)$ is the \ac{HIM} prediction of the \xmax{} variance for the $\langle \text{ln}(A) \rangle$ value found at this energy, and $\sigma^2_p$ is the \ac{HIM} \xmax{} variance prediction for proton; $b$ is a fit parameter.
This procedure is applied to the FD~\cite{Auger:2025InPrep}, SD~DNN~\cite{PierreAuger:2024nzw}, and HEAT~\cite{Bellido:2017cgf} results in \autoref{fig:XmaxMoments} using EPOS-LHC, Sibyll~2.3d and QGSJET-II.04.
The resulting moments of \lnA{} are shown in \autoref{fig:lnAMoments}.

\renewcommand\sidecaptionsep{2.2mm}
\sidecaptionvpos{figure}{c}

\begin{wrapfigure}{r}{0.6\textwidth}
  \centering
  \includegraphics[width=\linewidth]{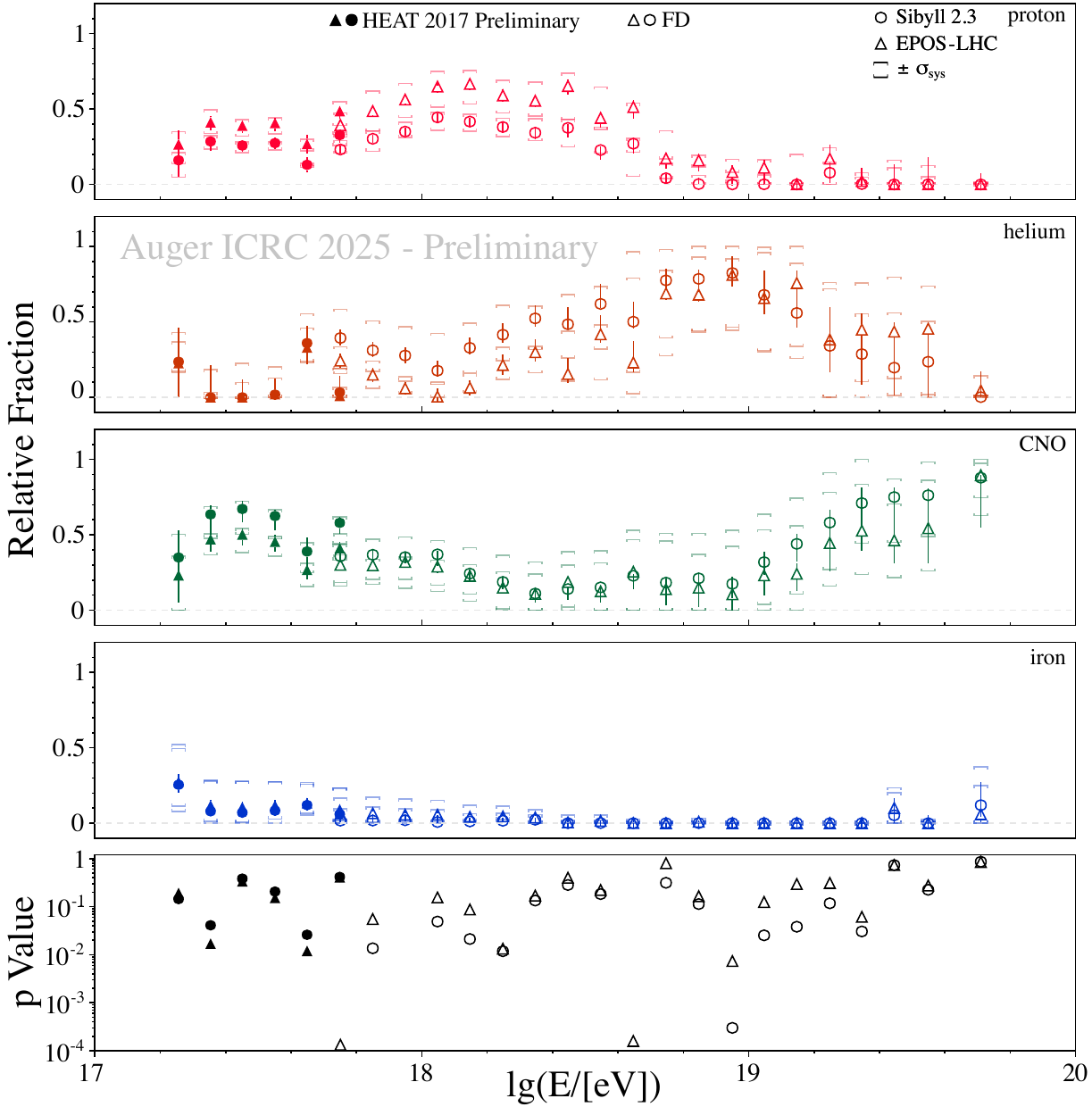}
  \caption{\ac{UHECR} fractional mass composition derived from FD~\cite{Auger:2025InPrep} and preliminary \ac{HEAT} data\cite{Bellido:2017cgf}. Estimated using Sibyll 2.3 (HEAT), Sibyll 2.3d, and EPOS-LHC. QGSJet-II.04 has been omitted as it produces unphysical results (see ~\autoref{fig:lnAMoments} for example). For details on the fits / analysis and fits with new models, see \cite{Auger:2025InPrep}.}
  \label{fig:Fractions}
\end{wrapfigure}
As can be seen, the first and second moments of \xmax{} and \lnA{} provide a clear summary of the overall UHECR composition.
However, they do not offer a clear picture of the individual contributions of distinct mass groups.
By generating templates of the \xmax{} distributions for proton, helium, nitrogen, and iron with \acp{HIM}, and then fitting a superposition of these templates to the measured \xmax{} distributions at each energy, estimates of the fractional abundances of each mass group can be extracted~\cite{PierreAuger:2014gko}.
This procedure has been applied to the Phase I hybrid FD data~\cite{Auger:2025InPrep} and preliminary HEAT data~\cite{Bellido:2017cgf} in \autoref{fig:Fractions}.

It should be emphasized that the above results depend on the \acp{HIM} used and that many new models and model tweaks are under development, which are expected to modify the \xmax{} scale and its variations meaningfully.
These, in turn, are expected to alter the details shown in \autoref{fig:lnAMoments} and \autoref{fig:Fractions} (see~\cite{Auger:2025InPrep}).
However, in addition to the above methods, there are other techniques, for example, the comparison of SD signals in \xmax{} (see \cite{ICRC25:Yushkov}) and close examinations of the energy evolution of \xmaxmu{} and \xmaxsigma{}~\cite{PierreAuger:2024flk}, which can inform trends without reliance on \acp{HIM} for interpretation.
By applying all these approaches to the accumulated data from Phase I of the Observatory, a clear and consistent picture of the mass composition of UHECRs emerges.
Above $10^{17.2}$\,eV, the arriving composition of the UHECR flux can be generally characterized by the following principal behaviors:
\begin{enumerate}[topsep=3pt]\setlength\itemsep{-0.1em}
  \item \textbf{Predominantly hadronic primaries:} Nearly all UHECR primaries are protons or heavier atomic nuclei, as no definitive observations of other particle types have yet been made~\cite{Gonzalez:2025dwv, Alvarez-Muniz:2025pir}.
  \item \textbf{Non-monotonic evolution with energy:} As energy increases, average nuclear mass decreases until reaching a minimum near 3\,EeV.
  Afterward, it rises steadily with energy (\autoref{fig:lnAMoments})~\cite{Auger:2025InPrep}. 
  \item \textbf{Structured evolution with energy:} The evolution of \xmax{} with energy shows structure with changes seen around 2\,EeV with the FD~\cite{PierreAuger:2014sui, Auger:2025InPrep} and 6.5, 11, and 31\,EeV with the SD (\cite{PierreAuger:2024flk, ICRC25:Universality}).
  \item \textbf{Changing composition purity:} For most energies, the composition flux of UHECR is mixed, often with multiple adjacent mass groups present simultaneously (\autoref{fig:Fractions} and \cite{ICRC25:Yushkov}).  
  Above a few EeV, the degree of mixing in the UHECR beam starts to decrease.
  Past $\sim10$\,EeV, the flux at any single energy becomes increasingly dominated by species within a narrow range of masses (\autoref{fig:lnAMoments} right and \cite{PierreAuger:2024nzw}).
\end{enumerate}

%% file: content/4-PAOAugerComparison.tex
\section*{$X$\textsubscript{max} in the Southern and Northern Skies}

\begin{figure}[!b]
    \centering
    \includegraphics[height=3.7cm]{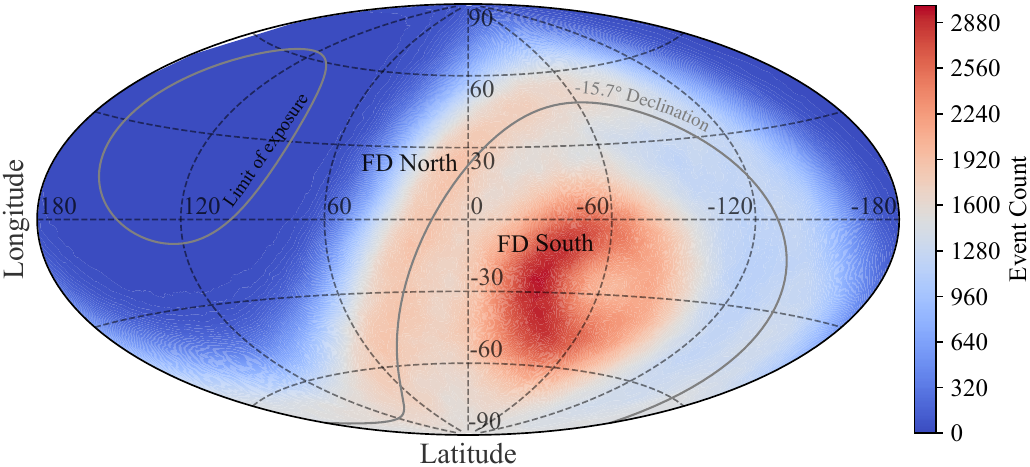}
    \includegraphics[height=3.7cm]{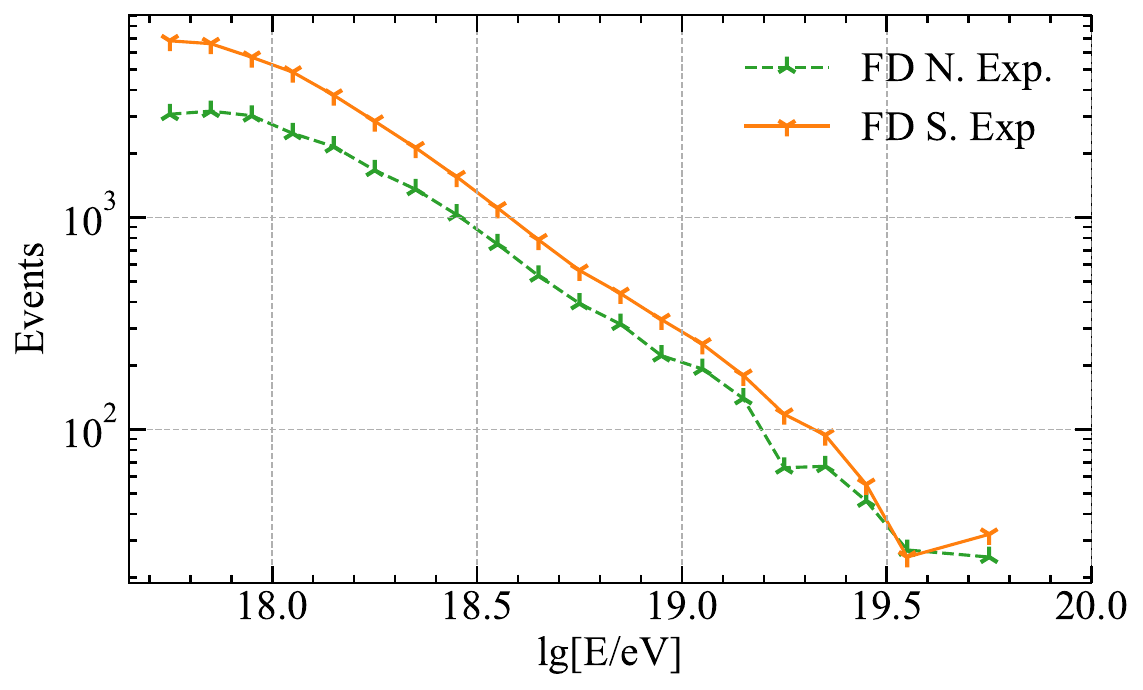}
    \caption{\footnotesize Northern and southern equatorial band definitions and counts. Left: FD Exposure with the $-15.7^\circ$ declination dividing line of the analysis split shown. Right: the relative counts of the analysis.}
    \label{fig:PAOTACompareMap}
\end{figure}

Due to the size of the Phase I hybrid dataset, it is now possible to split the sky into northern and southern regions in declination and carry out a preliminary independent \xmax{} analysis on both.
By comparing the \xmax{} measurements of these two sky regions, we can directly test whether \ac{UHECR} composition varies with declination within the Observatory’s exposure.
The split between the northern and southern regions is made at a declination of $-15.7^\circ$, which was chosen as it represents the southernmost extent of the FD hybrid composition sensitivity of Telescope Array~\cite{TelescopeArray:2018xyi}.
The exposure of the hybrid \xmax{} Phase I dataset of the Pierre Auger Observatory, with a line indicating the northern and southern data split, is shown on the left of \autoref{fig:PAOTACompareMap}. 
In contrast, the relative counts in each energy bin are shown on the right of \autoref{fig:PAOTACompareMap}.

\begin{SCfigure}[][!t]
    \centering
    \includegraphics[width=0.36\textwidth]{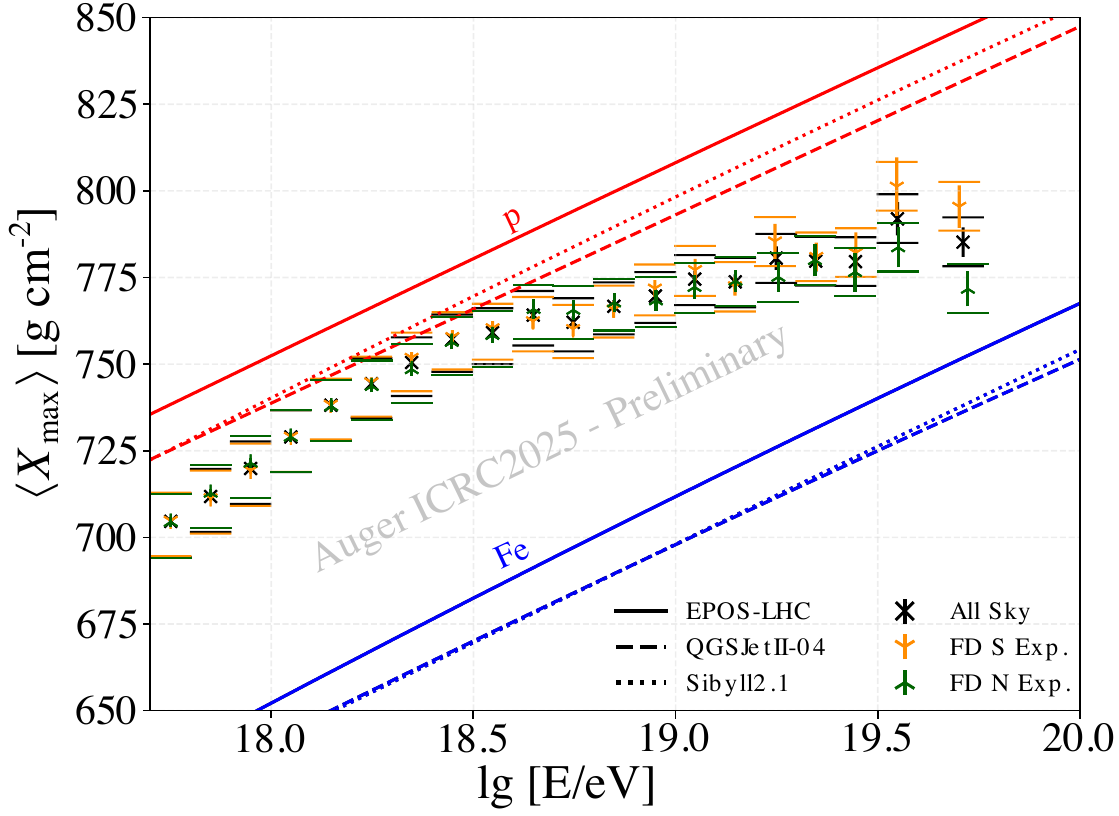}
    \includegraphics[width=0.36\textwidth]{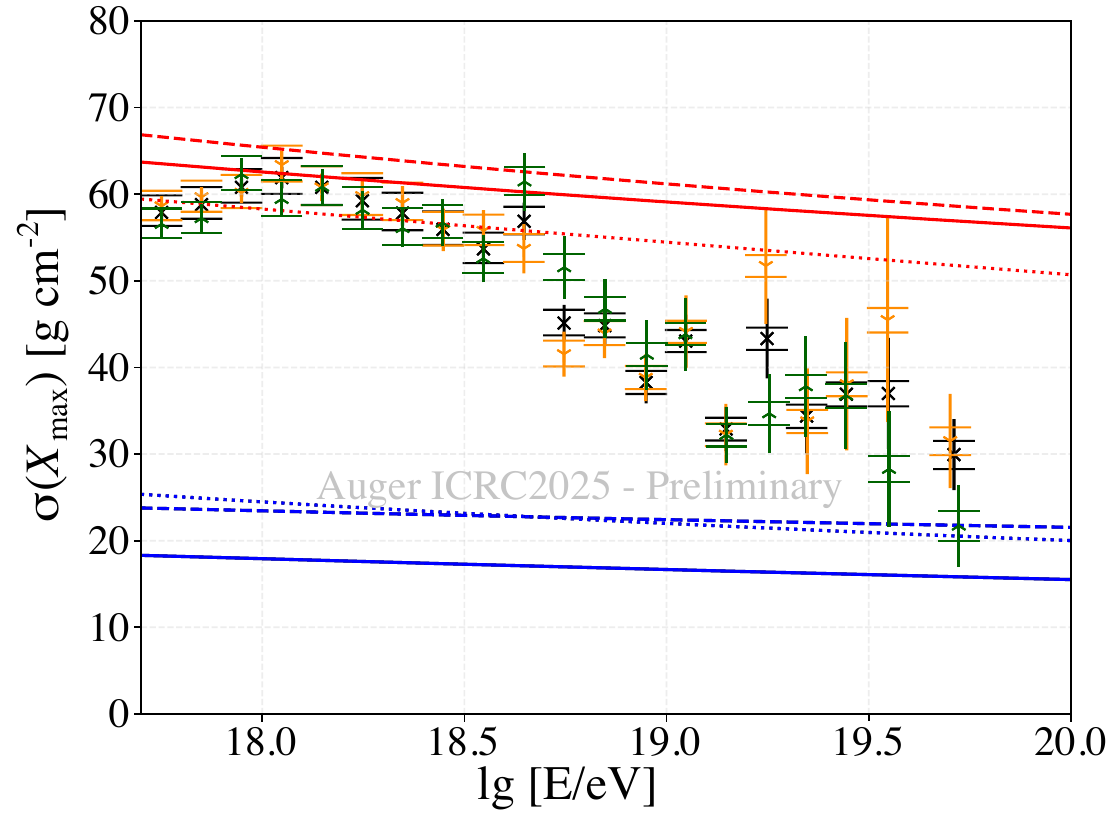}
    \caption{\footnotesize The first (left) and second (right) moments of the \xmax{} distributions from northern (green) and southern (orange) analysis regions. The black points represent the measurement over the full sky.}
    \label{fig:PAOTACompareMoments}
    \vspace{-3mm}
\end{SCfigure}

To compare the mass composition in these two regions, the same analysis as was carried out for the \ac{FD} hybrid data in \autoref{fig:XmaxMoments} was performed here.
However, for this analysis, separate investigations and corrections were carried out for each region to account for acceptance, reconstruction bias, resolution, and systematics, following the methods described in~\cite{PierreAuger:2021jlg}.
The differences in reconstruction bias, resolution, and other systematics were found to be small.
Because the distributions of shower zenith in the two regions differed, the \xmax{} dependent event acceptance also differed.
This acceptance was extracted using simulation, and its effects were corrected for using the $\Lambda_{\eta}$ method described in~\cite{PierreAuger:2014sui}.
Residual uncertainties between the two regions from this correction are less than 2\,\gcm{}.
The resulting moments of the \xmax{} distributions measured in each energy bin of the two regions are shown in \autoref{fig:PAOTACompareMoments}.

\renewcommand\sidecaptionsep{2.2mm}
\sidecaptionvpos{figure}{c}

\begin{SCfigure}[][!b]
    \vspace{-5mm}
    \centering
    \includegraphics[width=0.36\textwidth]{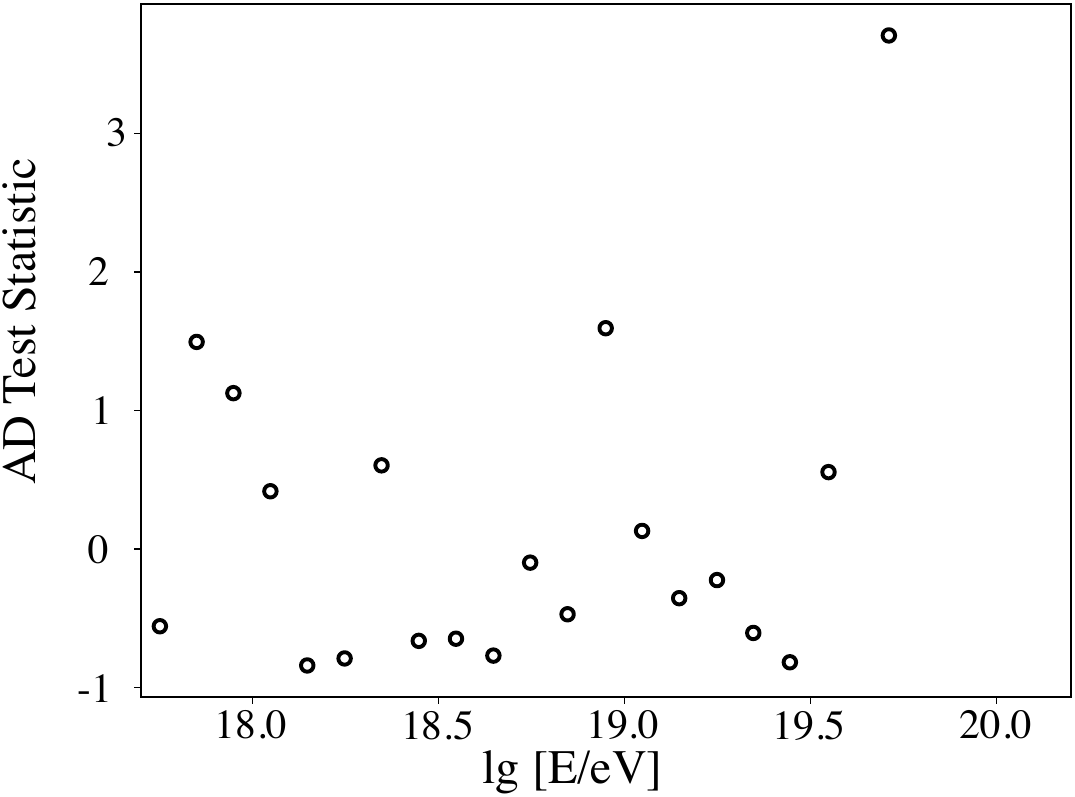}
    \includegraphics[width=0.36\textwidth]{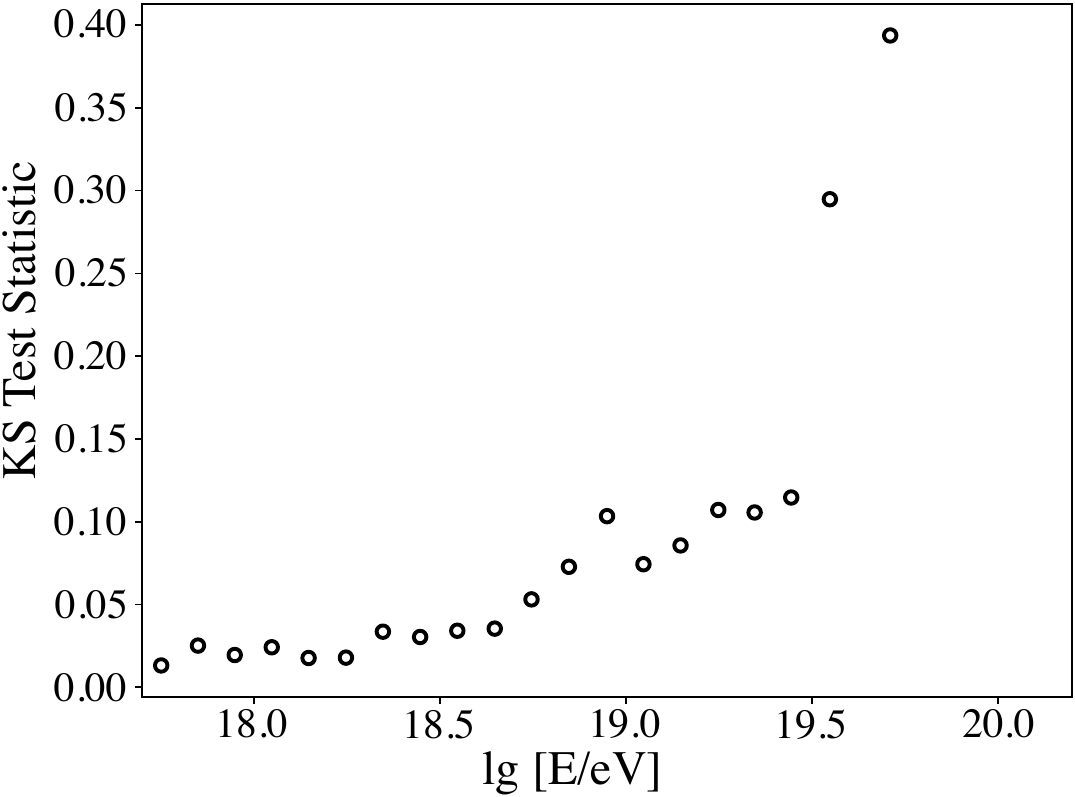}
    \caption{\footnotesize The results of KS (left) and AD (right) 2-sample tests between the \xmax{} distributions from northern and southern analysis regions.}
    \label{fig:PAOTACompareTests}
\end{SCfigure}

\autoref{fig:PAOTACompareMoments} left shows that there is excellent bin-to-bin agreement between the northern and southern skies for \xmaxmu{}.
In \autoref{fig:PAOTACompareMoments}, right, there is good statistical agreement for the majority of energies. 
Some differences in the moments do appear at high energies where statistics become poor.
To gauge whether any significant bin-by-bin differences occur between the two regions, the Kolmogorov-Smirnov~\cite{massey1951kolmogorov} (KS) and Anderson-Darling~\cite{scholz1987k} (AD) 2-sample tests are used to compare the two distributions in each energy bin in \autoref{fig:PAOTACompareTests}.

\sidecaptionvpos{figure}{c}

The results of the KS and AD tests indicate that the mass composition in the northern and southern regions is consistent with each other across the full energy range on a bin-by-bin basis.
Larger differences are seen in the last two energy bins with the KS (AD) test reporting a \ac{TS} of 0.29 (0.56) and 0.39 (3.72) for the second-to-last and last energy bins, respectively. 
For reference, the reduced $\chi^2$ for these bins are 1.6 and 3.0.
These KS (AD) values convert to local p-values of 0.016 (0.206) and 0.0003 (0.011) and global p-values of 0.322 (0.989) and 0.006 (0.20) for the second-to-last and last energy bins, respectively.
With the current statistics, the differences seen in these two bins are not globally significant.
This finding is in agreement with an earlier 2021 analysis, which found no evidence for a difference in the elongation rates and hence mass changes in published data taken in both hemispheres over multiple years~\cite{Watson:2021rfb}.
This result is also in full agreement with the work of the Pierre Auger Observatory/Telescope Array working group on mass composition, which found no tension in the mass measurements of the two Observatories~\cite{PierreAuger:2023yym}.
Further checks for differences between the northern and southern equatorial bands will be performed in the future with better statistics via SD and Phase II analyses.

%% file: content/5-Discussion.tex
\section*{Conclusion}

After nearly two decades of data taking, the Pierre Auger Observatory has amassed the most comprehensive dataset on UHECRs to date. 
Through the careful analysis of this data, the overall knowledge of the composition of arriving UHECRs has vastly improved. 
Within the limitations of our current Hadronic Interaction Models, we now know that \acp{UHECR} consist of many atomic nuclei species, ranging from protons to possibly up to iron nuclei. 
Their mass composition is mixed, evolves strongly with energy, and may have fine-grained features, similar to those observed in the spectrum. 
The ankle is not heavily dominated by protons, and flux trends toward heavy at the highest energies. 
To evaluate the possibility of a meaningful difference in mean composition between the northern and southern skies, the Phase I FD hybrid data were split at $-15.7^\circ$ declination.
When independent composition analyses were conducted on the two resulting equatorial bands, no significant differences in composition were found.

%% file: include/BackMatter.tex
\setlength{\intextsep}{12pt plus 2pt minus 2pt}
\setlength{\textfloatsep}{20pt plus 2pt minus 4pt}
\setlength{\abovecaptionskip}{10pt}
\setlength{\belowcaptionskip}{0pt}
\setlength{\abovedisplayskip}{12pt plus 3pt minus 9pt}
\setlength{\belowdisplayskip}{12pt plus 3pt minus 9pt}
\setlength{\abovedisplayshortskip}{0pt plus 3pt}
\setlength{\belowdisplayshortskip}{7pt plus 3pt minus 4pt}

\bibliographystyle{include/JHEPNoTitle.bst}
{\footnotesize
\bibliography{include/References.bib}{}
}

\clearpage
\section*{The Pierre Auger Collaboration}
\small
\noindent\hspace{0pt}\nobreak
\begin{wrapfigure}[8]{l}{0.11\linewidth}  
  \vspace{-5mm}
  \includegraphics[width=0.98\linewidth]{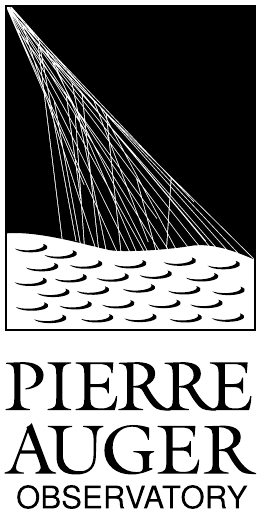}
\end{wrapfigure}
\noindent\input{include/latex_authorlist_authors}



\begin{center}
\rule{0.1\columnwidth}{0.5pt}
\raisebox{-0.4ex}{\scriptsize$\bullet$}
\rule{0.1\columnwidth}{0.5pt}
\end{center}

\vspace{-1ex}
\footnotesize
\input{include/latex_authorlist_institutions}

\vspace{-1ex}
\footnotesize
\input{include/acknowledgments}

%% file: include/latex_authorlist_authors.tex
A.~Abdul Halim$^{13}$,
P.~Abreu$^{70}$,
M.~Aglietta$^{53,51}$,
I.~Allekotte$^{1}$,
K.~Almeida Cheminant$^{78,77}$,
A.~Almela$^{7,12}$,
R.~Aloisio$^{44,45}$,
J.~Alvarez-Mu\~niz$^{76}$,
A.~Ambrosone$^{44}$,
J.~Ammerman Yebra$^{76}$
% ...

%% file: include/latex_authorlist_institutions.tex
\begin{description}[labelsep=0.2em,align=right,labelwidth=0.7em,labelindent=0em,leftmargin=2em,noitemsep,before={\renewcommand\makelabel[1]{##1 }}]
\item[$^{1}$] Centro At\'omico Bariloche and Instituto Balseiro (CNEA-UNCuyo-CONICET), San Carlos de Bariloche, Argentina
\item[$^{2}$] Departamento de F\'\i{}sica and Departamento de Ciencias de la Atm\'osfera y los Oc\'eanos, FCEyN, Universidad de Buenos Aires and CONICET, Buenos Aires, Argentina
\item[$^{3}$] IFLP, Universidad Nacional de La Plata and CONICET, La Plata, Argentina
\item[$^{4}$] Instituto de Astronom\'\i{}a y F\'\i{}sica del Espacio (IAFE, CONICET-UBA), Buenos Aires, Argentina
\item[$^{5}$] Instituto de F\'\i{}sica de Rosario (IFIR) -- CONICET/U.N.R.\ and Facultad de Ciencias Bioqu\'\i{}micas y Farmac\'euticas U.N.R., Rosario, Argentina
\item[$^{6}$] Instituto de Tecnolog\'\i{}as en Detecci\'on y Astropart\'\i{}culas (CNEA, CONICET, UNSAM), and Universidad Tecnol\'ogica Nacional -- Facultad Regional Mendoza (CONICET/CNEA), Mendoza, Argentina
\item[$^{7}$] Instituto de Tecnolog\'\i{}as en Detecci\'on y Astropart\'\i{}culas (CNEA, CONICET, UNSAM), Buenos Aires, Argentina
\item[$^{8}$] International Center of Advanced Studies and Instituto de Ciencias F\'\i{}sicas, ECyT-UNSAM and CONICET, Campus Miguelete -- San Mart\'\i{}n, Buenos Aires, Argentina
\item[$^{9}$] Laboratorio Atm\'osfera -- Departamento de Investigaciones en L\'aseres y sus Aplicaciones -- UNIDEF (CITEDEF-CONICET), Argentina
\item[$^{10}$] Observatorio Pierre Auger, Malarg\"ue, Argentina
\item[$^{11}$] Observatorio Pierre Auger and Comisi\'on Nacional de Energ\'\i{}a At\'omica, Malarg\"ue, Argentina
\item[$^{12}$] Universidad Tecnol\'ogica Nacional -- Facultad Regional Buenos Aires, Buenos Aires, Argentina
\item[$^{13}$] University of Adelaide, Adelaide, S.A., Australia
\item[$^{14}$] Universit\'e Libre de Bruxelles (ULB), Brussels, Belgium
\item[$^{15}$] Vrije Universiteit Brussels, Brussels, Belgium
\item[$^{16}$] Centro Brasileiro de Pesquisas Fisicas, Rio de Janeiro, RJ, Brazil
\item[$^{17}$] Centro Federal de Educa\c{c}\~ao Tecnol\'ogica Celso Suckow da Fonseca, Petropolis, Brazil
\item[$^{18}$] Instituto Federal de Educa\c{c}\~ao, Ci\^encia e Tecnologia do Rio de Janeiro (IFRJ), Brazil
\item[$^{19}$] Universidade de S\~ao Paulo, Escola de Engenharia de Lorena, Lorena, SP, Brazil
\item[$^{20}$] Universidade de S\~ao Paulo, Instituto de F\'\i{}sica de S\~ao Carlos, S\~ao Carlos, SP, Brazil
\item[$^{21}$] Universidade de S\~ao Paulo, Instituto de F\'\i{}sica, S\~ao Paulo, SP, Brazil
\item[$^{22}$] Universidade Estadual de Campinas (UNICAMP), IFGW, Campinas, SP, Brazil
\item[$^{23}$] Universidade Estadual de Feira de Santana, Feira de Santana, Brazil
\item[$^{24}$] Universidade Federal de Campina Grande, Centro de Ciencias e Tecnologia, Campina Grande, Brazil
\item[$^{25}$] Universidade Federal do ABC, Santo Andr\'e, SP, Brazil
\item[$^{26}$] Universidade Federal do Paran\'a, Setor Palotina, Palotina, Brazil
\item[$^{27}$] Universidade Federal do Rio de Janeiro, Instituto de F\'\i{}sica, Rio de Janeiro, RJ, Brazil
\item[$^{28}$] Universidad de Medell\'\i{}n, Medell\'\i{}n, Colombia
\item[$^{29}$] Universidad Industrial de Santander, Bucaramanga, Colombia
\item[$^{30}$] Charles University, Faculty of Mathematics and Physics, Institute of Particle and Nuclear Physics, Prague, Czech Republic
\item[$^{31}$] Institute of Physics of the Czech Academy of Sciences, Prague, Czech Republic
\item[$^{32}$] Palacky University, Olomouc, Czech Republic
\item[$^{33}$] CNRS/IN2P3, IJCLab, Universit\'e Paris-Saclay, Orsay, France
\item[$^{34}$] Laboratoire de Physique Nucl\'eaire et de Hautes Energies (LPNHE), Sorbonne Universit\'e, Universit\'e de Paris, CNRS-IN2P3, Paris, France
\item[$^{35}$] Univ.\ Grenoble Alpes, CNRS, Grenoble Institute of Engineering Univ.\ Grenoble Alpes, LPSC-IN2P3, 38000 Grenoble, France
\item[$^{36}$] Universit\'e Paris-Saclay, CNRS/IN2P3, IJCLab, Orsay, France
\item[$^{37}$] Bergische Universit\"at Wuppertal, Department of Physics, Wuppertal, Germany
\item[$^{38}$] Karlsruhe Institute of Technology (KIT), Institute for Experimental Particle Physics, Karlsruhe, Germany
\item[$^{39}$] Karlsruhe Institute of Technology (KIT), Institut f\"ur Prozessdatenverarbeitung und Elektronik, Karlsruhe, Germany
\item[$^{40}$] Karlsruhe Institute of Technology (KIT), Institute for Astroparticle Physics, Karlsruhe, Germany
\item[$^{41}$] RWTH Aachen University, III.\ Physikalisches Institut A, Aachen, Germany
\item[$^{42}$] Universit\"at Hamburg, II.\ Institut f\"ur Theoretische Physik, Hamburg, Germany
\item[$^{43}$] Universit\"at Siegen, Department Physik -- Experimentelle Teilchenphysik, Siegen, Germany
\item[$^{44}$] Gran Sasso Science Institute, L'Aquila, Italy
\item[$^{45}$] INFN Laboratori Nazionali del Gran Sasso, Assergi (L'Aquila), Italy
\item[$^{46}$] INFN, Sezione di Catania, Catania, Italy
\item[$^{47}$] INFN, Sezione di Lecce, Lecce, Italy
\item[$^{48}$] INFN, Sezione di Milano, Milano, Italy
\item[$^{49}$] INFN, Sezione di Napoli, Napoli, Italy
\item[$^{50}$] INFN, Sezione di Roma ``Tor Vergata'', Roma, Italy
\item[$^{51}$] INFN, Sezione di Torino, Torino, Italy
\item[$^{52}$] Istituto di Astrofisica Spaziale e Fisica Cosmica di Palermo (INAF), Palermo, Italy
\item[$^{53}$] Osservatorio Astrofisico di Torino (INAF), Torino, Italy
\item[$^{54}$] Politecnico di Milano, Dipartimento di Scienze e Tecnologie Aerospaziali , Milano, Italy
\item[$^{55}$] Universit\`a del Salento, Dipartimento di Matematica e Fisica ``E.\ De Giorgi'', Lecce, Italy
\item[$^{56}$] Universit\`a dell'Aquila, Dipartimento di Scienze Fisiche e Chimiche, L'Aquila, Italy
\item[$^{57}$] Universit\`a di Catania, Dipartimento di Fisica e Astronomia ``Ettore Majorana``, Catania, Italy
\item[$^{58}$] Universit\`a di Milano, Dipartimento di Fisica, Milano, Italy
\item[$^{59}$] Universit\`a di Napoli ``Federico II'', Dipartimento di Fisica ``Ettore Pancini'', Napoli, Italy
\item[$^{60}$] Universit\`a di Palermo, Dipartimento di Fisica e Chimica ''E.\ Segr\`e'', Palermo, Italy
\item[$^{61}$] Universit\`a di Roma ``Tor Vergata'', Dipartimento di Fisica, Roma, Italy
\item[$^{62}$] Universit\`a Torino, Dipartimento di Fisica, Torino, Italy
\item[$^{63}$] Benem\'erita Universidad Aut\'onoma de Puebla, Puebla, M\'exico
\item[$^{64}$] Unidad Profesional Interdisciplinaria en Ingenier\'\i{}a y Tecnolog\'\i{}as Avanzadas del Instituto Polit\'ecnico Nacional (UPIITA-IPN), M\'exico, D.F., M\'exico
\item[$^{65}$] Universidad Aut\'onoma de Chiapas, Tuxtla Guti\'errez, Chiapas, M\'exico
\item[$^{66}$] Universidad Michoacana de San Nicol\'as de Hidalgo, Morelia, Michoac\'an, M\'exico
\item[$^{67}$] Universidad Nacional Aut\'onoma de M\'exico, M\'exico, D.F., M\'exico
\item[$^{68}$] Institute of Nuclear Physics PAN, Krakow, Poland
\item[$^{69}$] University of \L{}\'od\'z, Faculty of High-Energy Astrophysics,\L{}\'od\'z, Poland
\item[$^{70}$] Laborat\'orio de Instrumenta\c{c}\~ao e F\'\i{}sica Experimental de Part\'\i{}culas -- LIP and Instituto Superior T\'ecnico -- IST, Universidade de Lisboa -- UL, Lisboa, Portugal
\item[$^{71}$] ``Horia Hulubei'' National Institute for Physics and Nuclear Engineering, Bucharest-Magurele, Romania
\item[$^{72}$] Institute of Space Science, Bucharest-Magurele, Romania
\item[$^{73}$] Center for Astrophysics and Cosmology (CAC), University of Nova Gorica, Nova Gorica, Slovenia
\item[$^{74}$] Experimental Particle Physics Department, J.\ Stefan Institute, Ljubljana, Slovenia
\item[$^{75}$] Universidad de Granada and C.A.F.P.E., Granada, Spain
\item[$^{76}$] Instituto Galego de F\'\i{}sica de Altas Enerx\'\i{}as (IGFAE), Universidade de Santiago de Compostela, Santiago de Compostela, Spain
\item[$^{77}$] IMAPP, Radboud University Nijmegen, Nijmegen, The Netherlands
\item[$^{78}$] Nationaal Instituut voor Kernfysica en Hoge Energie Fysica (NIKHEF), Science Park, Amsterdam, The Netherlands
\item[$^{79}$] Stichting Astronomisch Onderzoek in Nederland (ASTRON), Dwingeloo, The Netherlands
\item[$^{80}$] Universiteit van Amsterdam, Faculty of Science, Amsterdam, The Netherlands
\item[$^{81}$] Case Western Reserve University, Cleveland, OH, USA
\item[$^{82}$] Colorado School of Mines, Golden, CO, USA
\item[$^{83}$] Department of Physics and Astronomy, Lehman College, City University of New York, Bronx, NY, USA
\item[$^{84}$] Michigan Technological University, Houghton, MI, USA
\item[$^{85}$] New York University, New York, NY, USA
\item[$^{86}$] University of Chicago, Enrico Fermi Institute, Chicago, IL, USA
\item[$^{87}$] University of Delaware, Department of Physics and Astronomy, Bartol Research Institute, Newark, DE, USA
\item[] -----
\item[$^{a}$] Max-Planck-Institut f\"ur Radioastronomie, Bonn, Germany
\item[$^{b}$] also at Kapteyn Institute, University of Groningen, Groningen, The Netherlands
\item[$^{c}$] School of Physics and Astronomy, University of Leeds, Leeds, United Kingdom
\item[$^{d}$] Fermi National Accelerator Laboratory, Fermilab, Batavia, IL, USA
\item[$^{e}$] Pennsylvania State University, University Park, PA, USA
\item[$^{f}$] Colorado State University, Fort Collins, CO, USA
\item[$^{g}$] Louisiana State University, Baton Rouge, LA, USA
\item[$^{h}$] now at Graduate School of Science, Osaka Metropolitan University, Osaka, Japan
\item[$^{i}$] Institut universitaire de France (IUF), France
\item[$^{j}$] now at Technische Universit\"at Dortmund and Ruhr-Universit\"at Bochum, Dortmund and Bochum, Germany
\end{description}

%% file: include/acknowledgments.tex
\section*{Acknowledgments}

\begin{sloppypar}
The successful installation, commissioning, and operation of the Pierre
Auger Observatory would not have been possible without the strong
commitment and effort from the technical and administrative staff in
Malarg\"ue. We are very grateful to the following agencies and
organizations for financial support:
\end{sloppypar}

\begin{sloppypar}
Argentina -- Comisi\'on Nacional de Energ\'\i{}a At\'omica; Agencia Nacional de
Promoci\'on Cient\'\i{}fica y Tecnol\'ogica (ANPCyT); Consejo Nacional de
Investigaciones Cient\'\i{}ficas y T\'ecnicas (CONICET); Gobierno de la
Provincia de Mendoza; Municipalidad de Malarg\"ue; NDM Holdings and Valle
Las Le\~nas; in gratitude for their continuing cooperation over land
access; Australia -- the Australian Research Council; Belgium -- Fonds
de la Recherche Scientifique (FNRS); Research Foundation Flanders (FWO),
Marie Curie Action of the European Union Grant No.~101107047; Brazil --
Conselho Nacional de Desenvolvimento Cient\'\i{}fico e Tecnol\'ogico (CNPq);
Financiadora de Estudos e Projetos (FINEP); Funda\c{c}\~ao de Amparo \`a
Pesquisa do Estado de Rio de Janeiro (FAPERJ); S\~ao Paulo Research
Foundation (FAPESP) Grants No.~2019/10151-2, No.~2010/07359-6 and
No.~1999/05404-3; Minist\'erio da Ci\^encia, Tecnologia, Inova\c{c}\~oes e
Comunica\c{c}\~oes (MCTIC); Czech Republic -- GACR 24-13049S, CAS LQ100102401,
MEYS LM2023032, CZ.02.1.01/0.0/0.0/16{\textunderscore}013/0001402,
CZ.02.1.01/0.0/0.0/18{\textunderscore}046/0016010 and
CZ.02.1.01/0.0/0.0/17{\textunderscore}049/0008422 and CZ.02.01.01/00/22{\textunderscore}008/0004632;
France -- Centre de Calcul IN2P3/CNRS; Centre National de la Recherche
Scientifique (CNRS); Conseil R\'egional Ile-de-France; D\'epartement
Physique Nucl\'eaire et Corpusculaire (PNC-IN2P3/CNRS); D\'epartement
Sciences de l'Univers (SDU-INSU/CNRS); Institut Lagrange de Paris (ILP)
Grant No.~LABEX ANR-10-LABX-63 within the Investissements d'Avenir
Programme Grant No.~ANR-11-IDEX-0004-02; Germany -- Bundesministerium
f\"ur Bildung und Forschung (BMBF); Deutsche Forschungsgemeinschaft (DFG);
Finanzministerium Baden-W\"urttemberg; Helmholtz Alliance for
Astroparticle Physics (HAP); Helmholtz-Gemeinschaft Deutscher
Forschungszentren (HGF); Ministerium f\"ur Kultur und Wissenschaft des
Landes Nordrhein-Westfalen; Ministerium f\"ur Wissenschaft, Forschung und
Kunst des Landes Baden-W\"urttemberg; Italy -- Istituto Nazionale di
Fisica Nucleare (INFN); Istituto Nazionale di Astrofisica (INAF);
Ministero dell'Universit\`a e della Ricerca (MUR); CETEMPS Center of
Excellence; Ministero degli Affari Esteri (MAE), ICSC Centro Nazionale
di Ricerca in High Performance Computing, Big Data and Quantum
Computing, funded by European Union NextGenerationEU, reference code
CN{\textunderscore}00000013; M\'exico -- Consejo Nacional de Ciencia y Tecnolog\'\i{}a
(CONACYT) No.~167733; Universidad Nacional Aut\'onoma de M\'exico (UNAM);
PAPIIT DGAPA-UNAM; The Netherlands -- Ministry of Education, Culture and
Science; Netherlands Organisation for Scientific Research (NWO); Dutch
national e-infrastructure with the support of SURF Cooperative; Poland
-- Ministry of Education and Science, grants No.~DIR/WK/2018/11 and
2022/WK/12; National Science Centre, grants No.~2016/22/M/ST9/00198,
2016/23/B/ST9/01635, 2020/39/B/ST9/01398, and 2022/45/B/ST9/02163;
Portugal -- Portuguese national funds and FEDER funds within Programa
Operacional Factores de Competitividade through Funda\c{c}\~ao para a Ci\^encia
e a Tecnologia (COMPETE); Romania -- Ministry of Research, Innovation
and Digitization, CNCS-UEFISCDI, contract no.~30N/2023 under Romanian
National Core Program LAPLAS VII, grant no.~PN 23 21 01 02 and project
number PN-III-P1-1.1-TE-2021-0924/TE57/2022, within PNCDI III; Slovenia
-- Slovenian Research Agency, grants P1-0031, P1-0385, I0-0033, N1-0111;
Spain -- Ministerio de Ciencia e Innovaci\'on/Agencia Estatal de
Investigaci\'on (PID2019-105544GB-I00, PID2022-140510NB-I00 and
RYC2019-027017-I), Xunta de Galicia (CIGUS Network of Research Centers,
Consolidaci\'on 2021 GRC GI-2033, ED431C-2021/22 and ED431F-2022/15),
Junta de Andaluc\'\i{}a (SOMM17/6104/UGR and P18-FR-4314), and the European
Union (Marie Sklodowska-Curie 101065027 and ERDF); USA -- Department of
Energy, Contracts No.~DE-AC02-07CH11359, No.~DE-FR02-04ER41300,
No.~DE-FG02-99ER41107 and No.~DE-SC0011689; National Science Foundation,
Grant No.~0450696, and NSF-2013199; The Grainger Foundation; Marie
Curie-IRSES/EPLANET; European Particle Physics Latin American Network;
and UNESCO.
\end{sloppypar}